\documentstyle[12pt,epsf,epsfig]{article}

\begin{document}

\begin{center}

{\large \bf How to calibrate the polarization of a high-energy \\  proton
beam ? A theoretical prospect }

\vspace{5mm} 

C. Bourrely and \underline{J. Soffer}

\vspace{5mm} 

{\small\it  Centre de Physique Th\'eorique - CNRS - Luminy,\\
Case 907 F-13288 Marseille Cedex 9 - France\\
} 
\end{center}

\begin{center} ABSTRACT  

\vspace{5mm}

\begin{minipage}{130
mm} \small In view of the realistic possibility for operating
high-energy polarized proton beams in future collider machines, it is
highly desirable to propose for such beams, an absolute calibration
allowing to measure accurately their degree of polarization. We consider
more specifically one practical method based on $pp$ elastic scattering
near the forward direction and we discuss its limitations.
\end{minipage} 

\end{center}

Recent technical progresses based on the simple idea of the so called
Siberian Snake, allow to envisage very seriously nowdays, the use of
highly polarized proton beams of several hundreds GeV. This is one of
the key elements supporting the unique spin programme for $pp$
collisions, which will be undertaken in the near future at RHIC. It has
also motivated some detailed studies at DESY in order to decide whether
or not HERA could operate as a $ep$ collider with both electron and
proton beams polarized. If all this is demonstrated to be feasible,
still it remains necessary to provide a reliable method for measuring
the beam polarization at a rather accurate level, say a few percents.
Such a high-energy polarimeter must use a well-known polarization effect
and several possibilities have been considered, which will be discussed
in this workshop. Note that in the case of a high-energy polarized
electron beam, a powerful method based on Compton scattering, with a
laser beam[1], is used at SLAC and allows to determine the beam
polarization with a precision of less than $1\%$. For polarized proton
beams, one rather interesting candidate is the so-called Coulomb-Nuclear
Interference (CNI) polarimeter relying on an idea first suggested by J.
Schwinger[2] in 1948 and subsequently studied by other authors[3,4]. Let
us consider in $pp$ elastic scattering the single transverse spin
asymmetry $A_N$. This observable has a rather complicated expression in
terms of the five helicity amplitudes which describe $pp$ elastic
scattering[4], but as a first approximation one can write

\begin{equation}
A_N=-\frac{2Im(\phi_{nf}\phi_{f}^{\star})}{\vert \phi_{nf}\vert^2+2\vert
\phi_{f}\vert^2}\ ,
\end{equation}
where $\phi_{nf}$ is the dominant non-flip amplitude and $\phi_{f}$ is
the single-flip amplitude. However near the forward direction, say
$-t\sim 10^{-3} GeV^2$, both the nuclear and the Coulomb forces are
important and one assumes that $A_N$ is mainly due to the interference
between the hadronic non-flip $\phi_{nf}^N$ and the electromagnetic
single-flip $\phi_{f}^C$ amplitude, the first one being mainly
imaginary and the second one being real. This leads to an exact
expression for $A_N$, namely

\begin{equation}
A_N=\frac{\sqrt{-t}}{m_p}
\frac{(\mu_p-1)z}{1-(\rho-z)^2-\frac{t}{2m^2_p}(\mu_p-1)^2z^2}\ ,
\end{equation}
where $\mu_p$ is the magnetic moment of the proton and $m_p$ its mass.
Here $\rho$ denotes the real to imaginary ratio of $\phi^N_{nf}$ at
$t=0$ and $z=t_c/\vert t\vert$, $t_c=8\pi\alpha/\sigma_{tot}$, where
$\alpha$ is the fine structure constant and $\sigma_{tot}$ the $pp$
total cross section. This exact expression leads to the dotted curve in
Fig.1 at $45 GeV/c$, with a maximum value of about $4\%$ for $-t\simeq
3.10^{-3} GeV^2$. Note that it is almost energy independent, except
through the values of $\rho$ and $\sigma_{tot}$. The CNI kinematic
region has been investigated by the $E-704$ experiment at FNAL[5] which
has obtained some results for $A_N$ at $200 GeV/c$, consistent with the
prediction of eq.(2) (see Fig.3, open and close circles). However, due
to the lack of statistical precision, this data does not allow to check
the validity of the assumption $\phi^N_f=0$, we made to derive eq.(2).
This is a key issue to be sure that the measurement of $A_N$ in the CNI
region provides an absolute polarimeter. Unfortunately this is not the
case, as we will demonstrate now. Considering our lack of precise
knowledge on $\phi^N_f$, both experimentally and theoretically, and just
to illustrate our arguments, we make the simple assumption

\begin{equation}
\phi^N_f = \sqrt{-t}/m_p (b+ia)\phi^N_{nf}\ ,
\end{equation}
where $\sqrt{-t}/m_p$ is a required kinematical factor and $a,b$ are
two adjustable parameters[6]. In order to constrain the magnitude of $a$
and $b$, one should use the most accurate data on $A_N$, at the highest
available energy and in the lowest $t$ range, close to the CNI region.
This is the Serpukhov data[7] at $45 GeV/c$ shown in Fig.1, where in
addition to the case $a=b=0$ (dotted curve), we also display the case
$a=-0.02$, $b=0$ (solid curve). So this means that if $b=0$, the data
allow $-0.02<a<0$ and it is important to remark that the region of the
maximum of
$A_N$ remains stable around the value $(4.25\pm 0.25)\%$.
The situation changes drastically if we allow $b\not = 0$ as shown in
Fig.2, where the dashed curve corresponds to $a=0$, $b=-0.25$ and the
solid curve to $a=0.01$, $b=-0.25$ (dotted curve as before $a=b=0$). As a
result the data allows
$0<a<0.01$ and $-0.25<b<0$, but now near $-t=3.10^{-3} GeV^2$, the value
of $A_N$ changes a lot from about $4\%$ to $5.5\%$. This uncertainty can
be reduced if we modify $\phi^N_f$ by an {\it ad hoc} multiplicative
factor $[1-(t/0.45)^2]$, such that $A_N$ changes sign around $-t=0.45
GeV^2$, a feature of the data at several energies. This is shown in Fig.3
where the dashed curve corresponds to $a=0$, $b=-0.02$ and the solid
curve to $a=0.02$, $b=-0.02$ (dotted curve as before $a=b=0$).

Our final conclusion, which has also been reached in ref.[8], is that,
given the large uncertainty on $\phi^N_f$, one cannot presently consider
CNI as an absolute polarimeter. So to improve our knowledge on $\phi^N_f$
in the CNI region we suggest to perform once more the E704
experiment[5], with a fixed polarized target, to a much higher level of
accuracy in order to use it as a reliable calibration method.

\baselineskip 12pt

{\small\noindent[1]
M.L. Swartz, Proceedings of the 15$^{th}$ SLAC Summer Institute on
Particle Physics, 1987, SLAC report N$^o$ 328, p. 83.

\noindent[2] J. Schwinger, Phys. Rev. {\bf 73} (1948) 407.

\noindent[3] B.Z. Kopeliovich and I.I. Lapidus, Sov. J. Nucl. Phys. {\bf
19} (1974) 114.

\noindent[4]
N.H. Buttimore, E. Gotsman and E. Leader, Phys. Rev. {\bf D18} (1978)
694.

\noindent[5] N. Akchurin et al., 
 Phys.  Lett.{\bf 229B} (1989) 299; Phys. Rev. {\bf D48} (1993) 3026.

\noindent[6] For a similar approach see also N. Akchurin, N.H. Buttimore
and A. Penzo,  Phys. Rev. {\bf D51} (1995) 3944.

\noindent[7] A. Gaidot et al.,
 Phys.  Lett.{\bf 61B} (1976) 103.

\noindent[8] L. Trueman, preprint BNL 1996, hep-ph/9610316.
}

\vspace{2cm}
\begin{figure}[h]
\vbox{%
\hbox to\hsize{\epsfxsize=0.45\hsize
\epsffile[37 37 510 510]{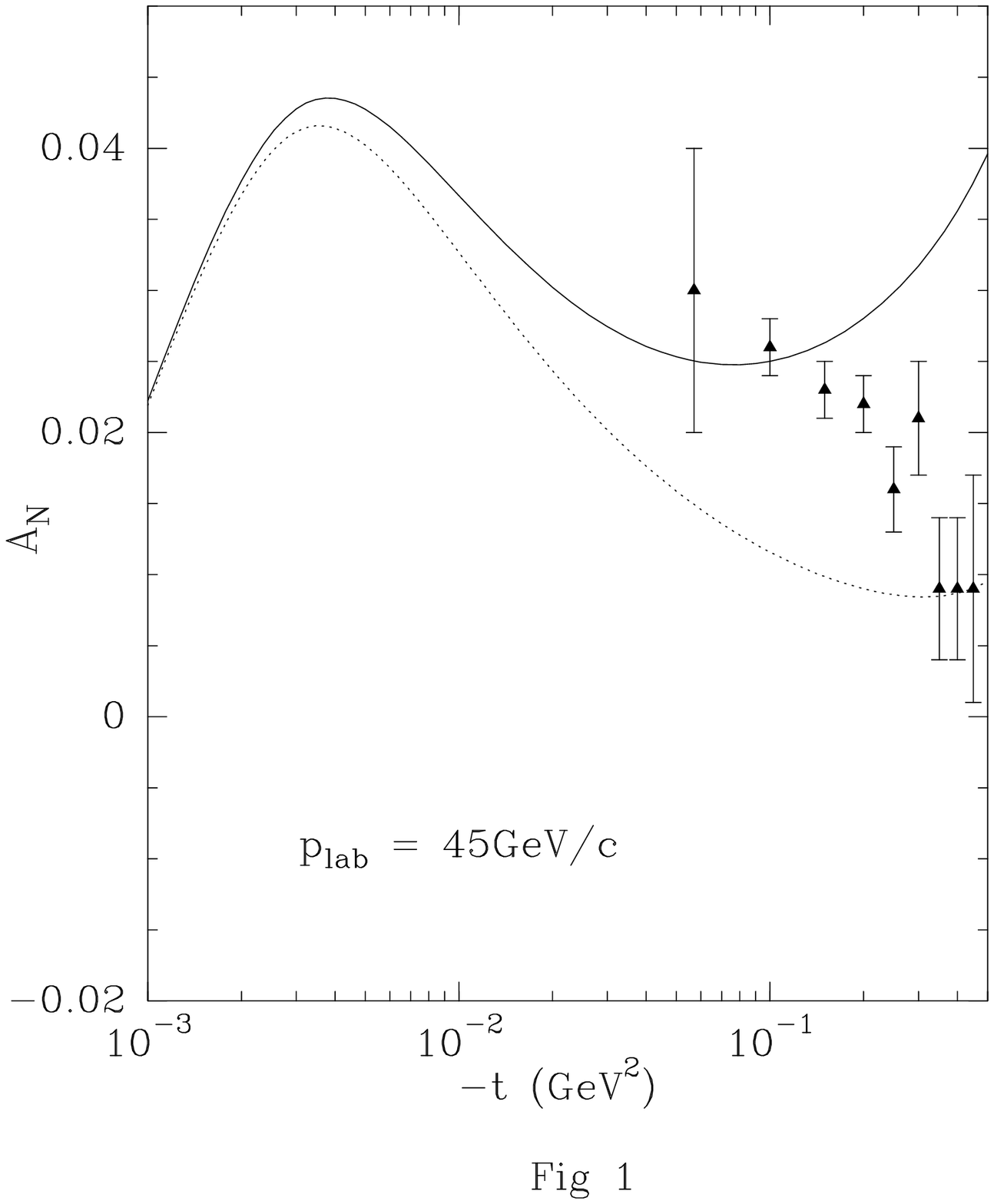}
\hfill
\epsfxsize=0.45\hsize\epsffile[37 37 510 510]{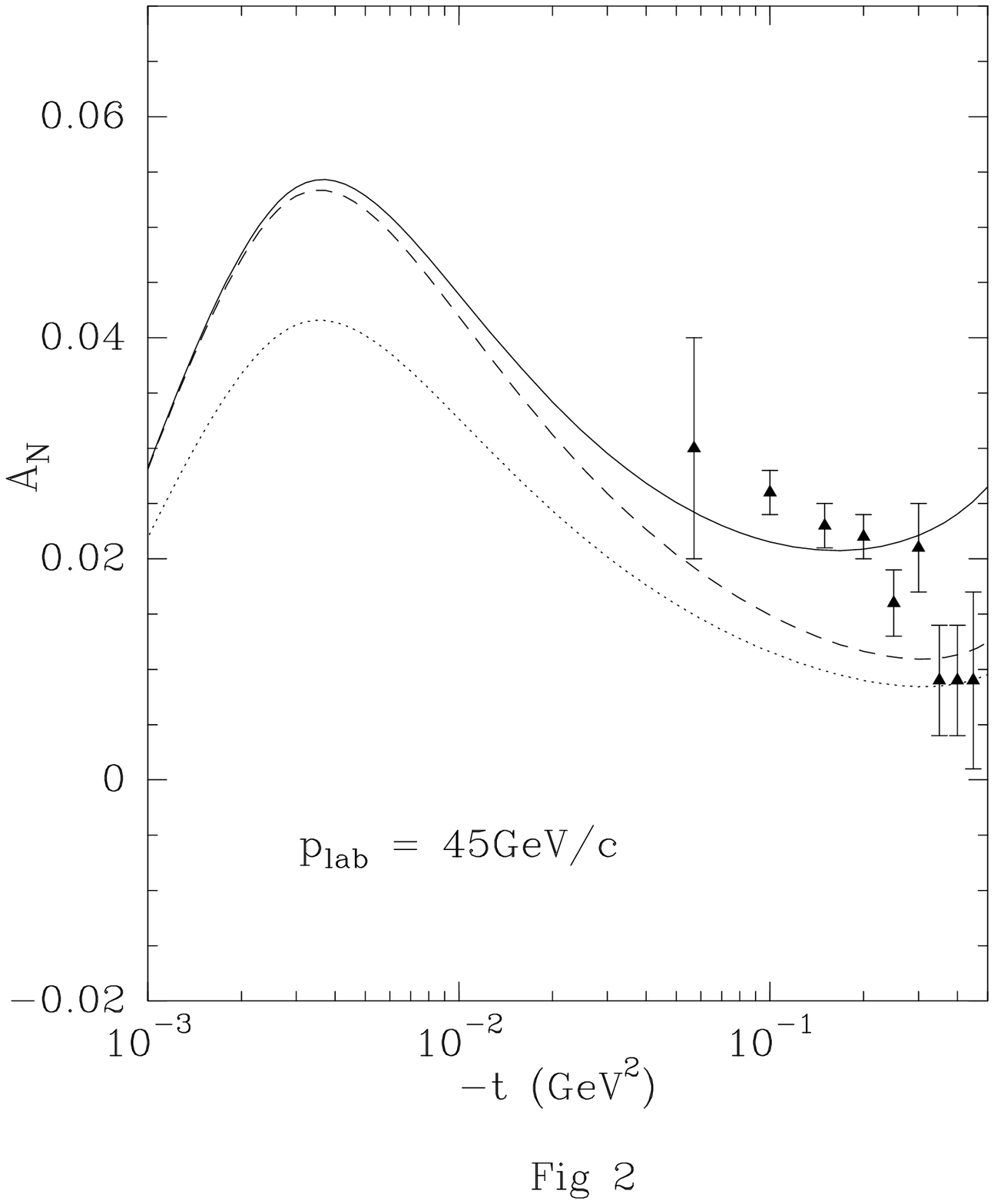}}
\vspace{1cm}
\hfill\hbox to\hsize{\epsfxsize=0.45\hsize
\epsffile[37 37 510 510]{fig3a.ps}
}\hfill}

\end{figure}

\end{document}